\documentclass[11pt]{article}
\usepackage{a4}
\usepackage{amsmath}
\usepackage{amsfonts}
\usepackage{latexsym}
\usepackage{epsfig}
\newcommand{\be}{\begin{equation}}
\newcommand{\ee}{\end{equation}}
\newcommand{\bea}{\begin{eqnarray}}
\newcommand{\eea}{\end{eqnarray}}

\renewcommand{\d}{\mathrm{d}}

\DeclareMathSymbol{\mg}{\mathrel}{symbols}{"1D}

\newcommand{\ga}{\alpha}
\newcommand{\gb}{\beta}
\renewcommand{\gg}{\gamma}

\renewcommand{\ge}{\epsilon}

\newcommand{\gf}{\phi}

\newcommand{\gm}{\mu}
\newcommand{\gn}{\nu}

\newcommand{\gr}{\rho}

\newcommand{\gs}{\sigma}

\newcommand{\go}{\omega}
\newcommand{\gz}{\zeta}
\newcommand{\gp}{\pi}

\newcommand{\gG}{\Gamma}
\newcommand{\gD}{\Delta}

\newcommand{\cP}{{\cal P}}

\newcommand{\cR}{{\cal R}}

\newcommand{\tr}{\text{tr}}

\newcommand{\ra}{\rightarrow}

\renewcommand{\Im}{\text{Im}}

\newcommand{\der}{\partial}

\newcommand{\labl}[1]{\label{#1}}
\newcommand{\half}{\frac 12 }
\newcommand{\shalf}{{\scriptstyle \half}}

\newcommand{\beq}{\begin{equation}}
\newcommand{\eeq}{\end{equation}}
\newcommand{\barr}{\begin{array}}
\newcommand{\earr}{\end{array}}
\newcommand{\equ}[1]{\begin{gather} #1 \end{gather}}
\newcommand{\equa}[1]{\begin{align} #1 \end{align}}

\newcommand{\enums}[1]{\begin{enumerate} #1 \end{enumerate}}

\newcommand{\non}{\nonumber}
\newcounter{oldcounter}

\newcommand{\Natr}{\mathbb{N}}
\newcommand{\Intr}{\mathbb{Z}}

\begin{document}

\begin{flushright} 
hep-th/0108185 
\end{flushright} 
\vskip 2 cm
\begin{center}
{\Large {\bf Dimensional regularization of a compact 
dimension.
}}
\\[0pt]
\bigskip {\large
{\bf Stefan Groot Nibbelink\footnote{
{{ {\ {\ {\ E-mail: nibblink@th.physik.uni-bonn.de}}}}}} }
\bigskip }\\[0pt]
\vspace{0.23cm}
{\it Physikalisches Institut der Universitat Bonn,} \\
{\it Nussallee 12, 53115 Bonn, Germany.}\\
\bigskip
\vspace{3.4cm} Abstract
\end{center}
{\small
An extension of dimensional regularization to the case of 
compact dimensions is presented. 
The procedure preserves the Kaluza-Klein tower 
structure, but has a regulator specific to the compact 
dimension.  Possible 5 and 4 dimensional divergent as well as manifest 
finite contributions of (one-loop) Feynman graphs can easily 
be identified in this scheme. 
}
\newpage

\section{Introduction}

It is well known that dimensional regularization in $4$ uncompact dimensions 
is a very powerful and convenient regularization scheme. The main idea 
is to obtain analytic expressions in the number of dimensions $D$ and then 
analytically continue it to $4$ dimensions. 
This scheme is universal in the sense that it can be applied to an arbitrary 
loop calculation of Feynman graphs in a quantum field theory. 
It respects all symmetries of the classical theory, except when this symmetry 
develops an anomaly at the quantum level. In addition, calculations with 
this procedure are simple and elegant since they rely on various properties 
of complex function theory.  

The central idea of this paper is to extend the dimensional regularization 
prescription such that it can also be applied to a space-time that has some 
compact dimensions in addition to the non-compact ones. We assume here a 
factorizable geometry $M^4 \times C$ where $M^4$ denotes 
the $4$ dimensional Minkowski space and $C$ is a compact manifold. 

Furthermore, we restrict our attention to the case where the compact 
manifold is one dimensional, i.e.\ typically a circle $S^1$, an orbicircle 
$S^1/\Intr_2$ or an orbifold $S^1/\Intr_2 \times \Intr_2'$. 
Models based on these types of orbifold have been the subject of 
study in the recent literature \cite{Barbieri,Delgado:98}. 
However the general regularization prescription that we present 
could in principle also be applied to other (more complicated) cases. 
Because of the compactness sums -- rather 
than integrals -- are obtained in momentum space. One would 
expect that in the limit of a large radius (small spacing of the 
momentum eigenvalues), a good approximation would be 
to replace the sum by the corresponding integral. In the approach 
described in this article we take this further and rewrite all sums as 
integrals. Since both sum and integrals are over momenta, the question 
which should be performed first should be irrelevant, in the 
regularization prescription we present this is indeed the case. 

Since the geometry of the manifold is a direct product $M^4 \times C$, 
we use two regulator parameters: $D_4$ denotes the complex extension 
of the dimension of the Minkowski space and $D_5$ denotes the 
complex extension of the single dimension of the compact manifold $C$. 
In addition we introduce two arbitrary 
renormalization scales $\gm_4$ and $\gm_5$. As we will see the 
appearance of $D_4$ and $D_5$ is very convenient because it allows us 
to trace whether a divergence has a 4 or 5 dimensional origin, depending 
on whether $D_5$, $D_4$ or $D_4 + D_5$ appear in a regulated 
expression. 

Dimensional regularization has been used before in the connection 
with compact manifolds. For example, in ref.\ 
\cite{Candelas,Sochichiu,DiClemente} 
it was combined with $\gz$-function regularization 
(see ref.\ \cite{Elizaldeetal} for a general review of this method) 
for the compact dimension. 
Hence there is only one regulator parameter for both the compact and 
non-compact dimensions. An attempt to have a separate regulator 
for compact dimensions is discussed in \cite{Contino}. The approach 
we discuss in this paper is more direct and makes a clear distinction 
between regulator effects and properties of the momentum spectrum 
due to the compact manifold. 

Furthermore, the regularization prescription can in principle be applied 
to any effective field theory description obtained by integrating out some 
additional dimension. Therefore, it can be used in situations more general 
than those considered in this paper. Just as an example, 
we mention that (loop) calculations with the mass spectra obtained 
in \cite{Gherghetta} for a warped geometry \cite{Rubakov,Randall} 
can be investigated with the method described here. (In fact, the 
asymptotic form of the KK masses are treated in this paper.)

In the following we first explain how after turning a sum into complex 
integral, the complex dimension $D_5$ can be introduced. The original  
properties of the sum are translated into the properties of a complex 
function that we call the ``pole function'' of the momentum spectrum. 
After a general description of such pole functions, we determine them 
for compactifications on a circle $S^1$ or orbicircle 
$S^1/\Intr_2$. Here we also discuss why one should 
sum the complete towers as is common practice in the ``Kaluza-Klein 
regularization''. Next we show that from the structure of 
those pole functions, we can identify the 4 and 5 dimensional divergences  
and finite contributions could be encountered in the calculation of a Feynman 
graph. A separate section is devoted to a discussion how fermions can be 
incorporated in this approach. 
We illustrate our method with the computation of the effective potential 
of the model presented in \cite{Barbieri}.

\section{Sum as $D_5$ dimensional integral} 
\labl{sumasint}

In this section we describe how a typical sum-integral, that arises from 
a Feynman graph in the 4 dimensional effective theory with KK towers, 
can be dimensionally regulated. Here we treat graphs with bosons only, 
however a large part of the discussion can directly be applied to graphs 
with fermions as well. Our discussion here is far from being complete, 
the main focus here is to convey the ideas behind the procedure. 

The Laplacian of the fifth dimension $\gD_5$ ($=\der_5^2$ for flat 
manifolds) has 
eigenfunctions $\gf_n$ with real positive eigenvalues $m_n^2$. 
Furthermore, let $f(p_4, p_5)$ be an arbitrary 
function of 4 dimensional momentum vector $p_4$ and 5 dimensional 
momentum $p_5$, where the latter can only take values $m_n$. 
For convenience,  we assume here that each eigenvalue $m_n$ 
has multiplicity $1$ and 
that $|m_n| \ra \infty$ if there are infinitely many eigenvalues. 
Furthermore we assume that $f(p_4, p_5)$ only depends on the 
combination $p_4^2 + p_5^2$. 
(These constrains are not really necessary as will become clear below. 
However, they make the discussion here more accessible.)
Only the infinite case ($n \ra \infty$) is discussed here, as it is easy to make a 
restriction to finite number of eigenvalues. 
We make the additional assumption that the function $f$ is meromorphic 
in $p_5$, but does not have poles on the real axis. This means that 
all poles $p_5 = X$ of $f(p_4, p_5)$ satisfy $|\Im \, X| \geq \ge$ for a given 
$\ge>0$ for any $p_4$. If the function $f$ does have poles on the 
real axis, then by a slight modification of this function they can 
be taken away from this axis by introducing an infrared regulator. 
In particular, this may be needed if one considers massless particles 
in the 5 dimensional theory. 

After these generalities, we want to compute the sum-integral 
\equ{
\int \d^4 p_4\, 
\sum_{n\in \Natr} \, f(p_4, m_n)  
\labl{beginsum}
} 
using dimensional regularization for both the four dimensional integral 
and for the sum. (All momentum integrals are Euclidean; the 
Wick rotation from Minkowski space is assumed to be performed.)  
To turn the sum over the eigenvalues into an 
integral, we define the pole function as the meromorphic function 
$\cP(p_5)$ with the properties:
\enums{
\item its set of poles is the set eigenvalues $\{m_n\ |\  n \in \Natr\}$, 
\item the residue at each of these poles is 1, 
\item for $p_5 \ra \infty$ with $\pm \Im\, p_5 > \ge$ its goes to an 
imaginary   constant $ \cP(p_5)  \ra \mp i \, r$.
}
Existence and uniqueness of $\cP$ follows from Mittag-Leffler's 
theorem which is closely related to the Weierstrass' product 
theorem, see for example \cite{Whittaker,Markushevich}. The first two 
conditions only determine the meromorphic function up to an arbitrary 
holomorphic function. The third condition fixes this function. 
The constant $r$ measures the size of the fifth dimension. In the examples 
of the circle and orbicircle  discussed in the next section 
we obtain precise relations between this 
quantity $r$ and their radius $R$. 
 
Let us first consider the situation where the infinite sum is convergent: 
to be more precise we assume that $C, \ga >0$ exist (which may be $p_4$ 
dependent) such that $|f(p_4, p_5)| \leq C |p_5|^{-1-\ga}$.  
Around any eigenvalue $m_n$ we can consider a contour of the form of 
a box $\Box_{(n, \ge)}$ with height $2\ge$ symmetric around the real axis. 
These boxes can be assumed be infinitely close together, their union is 
denoted by $\rightleftharpoons \; \equiv \cup_n \Box_{(n, \ge)}$. 
This contour goes infinitely close around the real axis.
Using standard contour integration the sum is rewritten 
as an integral \cite{Arkani,Mirabelli}. 
This integral can be rewritten as contour integral $\ominus$ over 
the upper and lower half plane with opposite orientation 
(anti-clockwise) to the $\rightleftharpoons$ contour. Here we have 
used that the arc contours at infinity vanish because of the bound 
on the function $f$ given above. 
\equa{
\sum_{n \in \Natr} f (p_4, m_n) = &\  
\sum_{n \in \Natr} 
\frac {-1}{2\gp i} \int_{\Box_{(n, \ge)}} \d p_5\, 
\cP(p_5) f(p_4, p_5) 
=
\labl{convsum}
\\
= &\ \frac {-1}{2\gp i} \int_{\rightleftharpoons} \d p_5\, 
\cP(p_5) f(p_4, p_5) 
= 
\frac {1}{2\gp i} \int_{\ominus} \d p_5 \, 
\cP(p_5) f(p_4, p_5).
\non
}
The figure below gives a schematic picture of this situation in the 
complex $p_5$-plane: 
\begin{center}
\epsfig{file=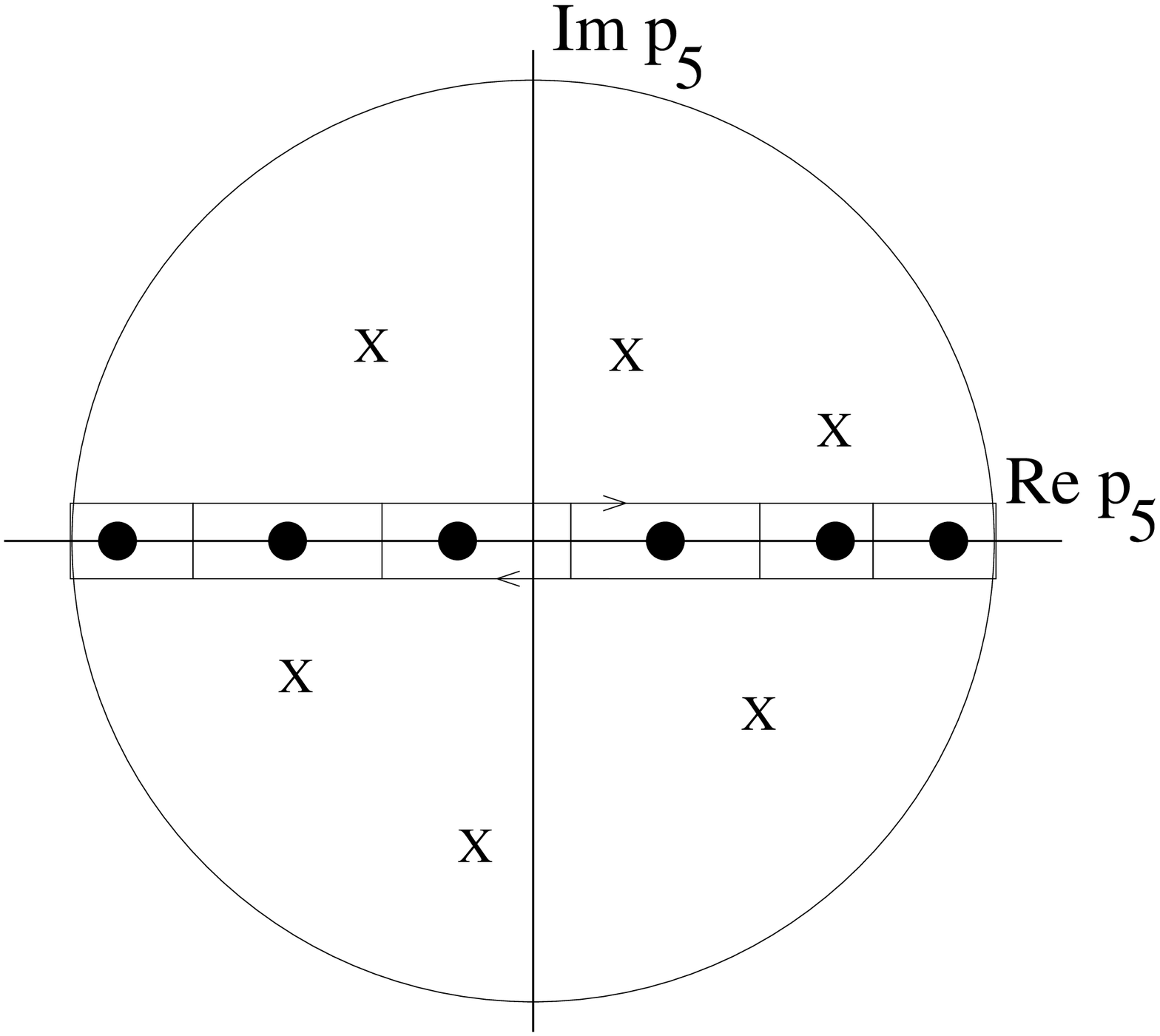, width=4cm, angle =0}
\end{center}
The symbols ``$\mbox{X}$'' denote the positions of poles $f$, and 
the dots $\bullet$ denote the real values which are the eigenvalues $m_n$; 
the poles of $\cP$. The contour orientations are described above. 

Now we want to set up a method that can handle a (divergent) 
sum-integral following the idea of replacing the sum by the 
contour integral $\ominus$. 
The dimensional regularization procedure is defined by the 
following replacement:
\equa{
\int \d^4 p_4\, 
\sum_{n\in \Natr} \, f(p_4, m_n)  \ra & \ 
\frac {1}{2\gp i}\,  \int_\ominus \d^{D_5} p_5 \int \d^{D_4} p_4\, 
\cP(p_5) f(p_4, p_5) \equiv 
\labl{dimreg} \\ 
& \ 
\frac {1}{2\gp i} \int_{\ominus} \d p_5 \int_0^\infty \d p_4 
\, \cR_{4}( p_4) \cR_{5}(p_5) \, 
\cP(p_5) \, f(p_4, p_5).
\non 
}
where we have introduced the regulator functions $\cR_{4}(p_4)$ and 
$\cR_5(p_5)$ with regulators $D_4$ and $D_5$ 
for the 4 dimensional and 5 dimensional integrations, given by
\equ{
\cR_{4}(p_4) =  
\frac {2 \gp^{\half (D_4)}}{\gG(\half D_4)}\, p_4^3 \, 
\Bigl( \frac {p_4}{\gm_4} \Bigr)^{D_4 -4}, 
\qquad 
\cR_{5}(p_5) =  
\frac {\gp^{\half (D_5)}}{\gG(\half D_5)}\, 
\Bigl( \frac {p_5}{\gm_5} \Bigr)^{D_5 -1}.
\labl{regufunctions}
}
Here we have introduced two (arbitrary) renormalization scales 
$\gm_4$ and $\gm_5$. 
The regulator function $\cR_{4}$ is the standard function for 
dimensional regularization of 4 non-compact dimensions \cite{tHooft}. 
The motivations for the regulator function $\cR_{5}$ are the following: 
it should reduce to unit when $D_5 \ra 1$. The straight line 
$-\infty < p_5 < \infty$ becomes a sphere with infinite radius in more 
(integer) dimensions, therefore one expects the usual volume factor  
${\gp^{\half (D_5)}}/{\gG(\half D_5)}$. To define the regulator 
function $\cR_5$ we need to introduce the complex logarithm which 
has a branch cut. Because the contour $\ominus$ does not contain the 
real line, it is convenient to take it along the negative real axis. 

For the convergent sum \eqref{convsum}, that serves at the inspiration 
for our regularization procedure, it is irrelevant in which order the 
sum is performed. In more mathematical words: for any bijection  
$P: \Natr \ra \Natr$, we have that 
\equ{
\sum_{n \in \Natr} f(p_4, m_n) = \sum_{n \in \Natr} f(p_4, m_{P(n)}). 
}
In physics such bijections can be interpreted as symmetries of the 
spectrum. An important property of the regularization prescription 
in \eqref{dimreg} is that it preserve such symmetries. The reason 
for this is that the sums are determined by the pole functions $\cP$ 
and $\cP_P$, where the latter is determined by the set of poles 
$\{m_{P(n)} \ |\ n \in \Natr\}$. But since this set and the original set  
$\{m_n \ |\ n \in \Natr\}$ that determines $\cP$ are identical, it follows 
that $\cP_P = \cP$. Note that such bijections cannot be interpreted 
on the level of the complex integration variable $p_5$.

This procedure regulates the sum and the integral separately and 
the prescription is independent of the mass structure of the KK 
theory that is encoded in the pole function $\cP$. The properties 
of the pole function determine the structure of possible divergences: 
in the next section we make a distinction between 5 and 4 dimensional 
divergences and finite contributions. The pole function $\cP$ 
contains information about the physical momentum spectrum; 
the regulator functions are just to make standard manipulations 
with divergent sum-integrals meaningful.

\section{Properties of the pole function}
\labl{proppolefunction}

Up to now the discussion was general in the sense that it did not specify 
the structure of the compact manifold.  We now restrict the 
attention to the circle $S^1$ and the orbicircle $S^1/\Intr_2$, to 
be able to exemplify important properties of the pole function. 

The mode functions are 
\(
\gf_n(x_5) = e^{i n x_5/R}, ~ n \in \Intr,
\)
for $S^1$ and 
\(
\gf_n^+(x_5) = \cos \frac {nx_5}R, ~ n \geq 0,
\)
and
\(
\gf_n^-(x_5) = \sin \frac{nx_5}R, ~ n > 0
\)
for $S^1/\Intr_2$. The corresponding eigenvalue momentum can be read 
off easily. It is straightforward to check that the pole function for 
$S^1$ is given by $\cP_0(p_5)$, with 
\equ{
\cP_\go(p_5) = \frac {\gp R}{\tan \gp R(p_5 - \go)}.
\labl{polefunSgo}
}
This more general form of the pole function $\cP_\go(p_5)$ 
describes a momentum spectrum that is shifted by $\go$. 
This form we use in the calculation of the effective potential discussed in 
section \ref{EffPot}. 

As the pole function is determined by the KK momentum, it is clear 
that it reflects the symmetries of the KK towers and that these symmetries 
are not broken by this regularization procedure. In the case of a 
compactification on a circle, the shifts over $2\gp R$ leave the manifold 
invariant, hence the momentum in the 5th dimension is quantized to 
be elements of $\Intr/R$. The set of integers are invariant under shifts by 
any integer. These shifts are symmetries of the pole function $\cP_0$ and 
hence respected by the dimensional regularization procedure applied to 
the compact dimension, since they define bijections $P_m(n) = n +m$ with 
$m$ a given integer. 

For the pole functions of the orbicircle a slight complication arises 
because the integer $n$ is either non-negative or positive. However, 
since we assumed that $f$ is a function of $p_5^2$ it is symmetric 
under the interchange of $n \ra -n$, therefore may take the sum over 
$n \in \Intr$ instead and divide by 2. Only for $n = 0$ we have to be 
a bit more careful: if $n = 0$ is included in the sum, $\half \frac 1{p_5}$ 
has to be added to give the pole $\frac 1{p_5}$ residue $1$, while if 
$n = 0$ is not included $\half \frac 1{p_5}$ has to be subtracted. 
This gives 
\equ{
\cP^{\pm}(p_5) = \half 
\left(
\pm \frac 1{p_5} + \cP_0(p_5) 
\right) 
= \half 
\left(
\pm \frac 1{p_5} + \frac {\gp R}{\tan \gp R p_5} 
\right). 
\labl{polefunS2}
}

The pole function $\cP_\go$ has important properties that can be 
used in identifying the structure of divergences that can occur. 
We define the remainder $\gr_{\go\, \pm}$ of $\cP_\go$ by 
\equ{
\cP_\go(p_5) = \mp i \gp R + \gr_{\go\, \pm}(p_5).
\labl{remainder}
}
It is not difficult to show that the remainder is bounded by  
\equ{
\gp^2 R^2 \, | 1 \mp \tanh x |^2 
\leq | \gr_{\go\, \pm}(p_5)|^2  \leq 
\gp^2 R^2\, \frac { | 1 \mp \tanh x |^2 }{ \tanh^2 x},
\labl{boundremainder}
}
with $x = \gp R \, \Im\, p_5$. From these inequalities the following 
important bounds can be derived.  Using the assumption that 
there is an $\ge > 0$ such that $|x| =  \gp R \, |\Im\, p_5| > \ge$ 
one obtains 
\equ{
\gp^2 R^2\, e^{- 4 |x|} 
\leq  | \gr_{\go\, \pm}(p_5)|^2  \leq 
\gp^2 R^2\,  \frac 1{\sinh^2 \ge} \, e^{- 4 |x|},
\labl{boundrem}
}
where for $ x > \ge$ the bounds apply to $\gr_{\go\, +}$ and 
for $ -x > \ge$ the bound apply to $\gr_{\go\, -}$, respectively. 

With this result we can now identify the different possible divergences 
that can occur. As an example, we consider the sum-integration of 
the function $f$ for the even mode functions $\gf^+_n$ on the orbicircle
\equ{
I = \int\d^4 p_4 \sum_{n\geq 0}  f(p_4, \frac nR) 
\ra 
\frac {1}{2\gp i}  \int \d^{D_4} p_4 \, \int_{\ominus} \d^{D_5} p_5 \, 
\cP^+(p_5) f(p_4, p_5) . 
}
The dimensionally regulated sum-integral now naturally splits into three 
parts $I = I_{5\,D} + I_{4\,D} + I_{finite}$. 
The constant parts $\pm \half i \gp R$ of the pole function 
$\cP^{+}$, see \eqref{remainder}, 
give the contributions 
\equa{
I_{5\,D} = 
 \frac {1}{2 \gp i}\, &
\frac {2 \gp^{\half (D_4 + D_5)}}{\gG(\half D_4)\gG(\half D_5)} 
\frac 1{\gm_4^{D_4-4} \gm_5^{D_5-1}} 
 \int_0^\infty \d p_4 \, p_4^{D_4-1}\,  \int_{-\infty}^\infty \d p_5 
\\ & \
\left\{
p_5^{D_5 -1} f( p_4, p_5) \Bigl(-\frac{i \gp R}2 \Bigr) + 
e^{i\gp} ( e^{i \gp} p_5)^{D_5 -1} 
f(p_4, e^{i\gp} p_5) \Bigl( \frac {i \gp R}2 \Bigr) 
\right\}. 
\non
}
Here we used that it is possible  to take the real part of  $D_5$ 
small enough such that the arc contours at infinity do not contribute.
Since in $f$ only $p_5^2$ appears, the factor $e^{i\gp}$ is 
irrelevant for the insertion of $p_5$ in $f$. We can restrict 
the integration over $p_5$ to $0 < p_5 < \infty$ by including an 
additional factor $2$. Since the function $f$ only depends on 
$p_4$ and $p_5$ via the combination $p_4^2 + p_5^2$, we can 
employ a change of variables 
$p_5 = p t^\half, ~ p_4 = p (1 - t)^\half$ with $0 < p < \infty$ 
and $0 < t < 1$.  Combining this with the standard result 
\equ{
\int_0^1 \d t\, t^{\ga -1} (1 - t)^{\gb -1} = 
\frac {\gG(\ga)\gG(\gb)}{\gG(\ga + \gb)}, 
}
the divergent part finally reads  
\equ{
I_{5\,D} = - \frac { 1 + e^{i \gp( D_5 -1)}}{4} \, 
 \frac{R}{\gm_4^{D_4-4} \gm_5^{D_5-1}} \, 
\frac {2 \gp^{\half (D_4 + D_5)}}{\gG(\shalf (D_4+ D_5))} 
\int_0^\infty \d p \, p^{D_4 + D_5 -1} f(p).
}
Here we denote with $p = (p_4, p_5)$ the full 5 dimensional 
momentum. This part behaves identically to a purely 5 dimensional 
integral since depends on $D_4 + D_5$ only. Therefore, its possible 
divergence has a 5 dimensional character. Notice that this calculation 
shows that the sum-integral can also be regulated using either $D_4$ 
or $D_5$. Actually, in dimensional regularization this contribution 
is regulated to be finite even if the regulators are taken away. However, 
this is just an artifact of dimensional regularization: if it is applied to 
a Minkowski space-time of odd dimension, all regulated integrals are 
finite \cite{Candelas}. 

Next we investigate the contribution due to the term $\half \frac 1{p_5}$ 
in the pole function $\cP^+$. As this term is a pole function that just 
gives only a contribution at $p_5 = 0$ there is no need to regulate the 
sum here anymore. It is similar to a contribution of a single particle. 
Therefore, we set $D_5 = 1$, perform the integration over contour 
$\rightleftharpoons$, and obtain a purely 4 dimensional integral
\equ{
I_{4\,D} 
= \half \, 
 \frac{1}{\gm_4^{D_4-4}} \, 
\frac {2 \gp^{\half D_4}}{\gG(D_4)} 
\int \d p_4 \, p_4^{D_4  -1} f(p_4, 0).
}

Finally, due to $\half\gr_{0\,\pm}$ in the pole function \eqref{remainder}
we obtain the contribution 
\equ{
I_{finite} = 
\frac {1}{2\gp i}  \int \d^{D_4} p_4 \, \int_{\ominus} \d^{D_5} p_5 \, 
\half \gr_{0\, \pm}(p_5) f(p_4, p_5), 
}
where we sum over $\pm$. 
The poles $p_{5\,\pm i} = \sqrt{(p_4 - k_i)^2 + m_i^2}$ of the function 
$f(p_4, p_5)$ in $p_5$ are labeled by the finite number of different pole 
pairs. Possible external 4 dimensional momenta are 
denoted by $k_i$ and $m_i^2$ are some masses that appear in the 
propagators. The contour around $\ominus$  contains all these poles. 
Therefore we get contributions with factors $\gr_{0\,\pm}(p_{5\, \pm i})$. 
We can conclude that these contributions are finite: integrating the 
bounds on the integrants \eqref{boundremainder} give convergent results. 
As we only get a finite number of pole contributions, 
we conclude that this contribution $I_{finite}$ is indeed finite. This means 
we can take $D_4 = 4$ and $D_5 =1$ to calculate this contribution.

\section{Inclusion of fermions}
\labl{inclferm}

The procedure we described so far applies to bosons only. The 
extension of this formulation with fermions is very similar to the 
situation in the well-known 4 dimensional case. We first review  
this briefly and then turn to the generalization to 5 
dimensions of which one is compact. 

There are three basic ways of regulating Clifford algebra properties
with the extension of the momentum integrals to arbitrary 
complex dimensions \cite{Buras}: 
naive dimensional regularization, 
dimensional regularization and dimensional reduction. 
The naive dimensional regularization scheme just introduces 
additional gamma matrices that anti-commute with each other 
and the 4 dimensional gamma matrices including $\gg^5$. 

In dimensional regularization the 4 dimensional Clifford algebra is 
extended to $D$ gamma matrices $\gg^\gm$ of which the 
first 4 are the original gamma matrices in 4 dimensions. They  
anti-commute to the generalized metric in $D$ dimensions. 
However, the chirality operator $\gg^5$ has a special role 
\cite{tHooft}: 
although it anti-commutes with the original gamma matrices,  
it commutes with the additional ones. This treatment is fully 
consistent  \cite{Breitenlohner}  and produces the axial-anomaly. 

In dimensional reduction the Clifford algebra and the spinor 
traces are worked out in 4 dimensions. 
The remaining momentum integrals are extended 
to $D$ dimensions is the standard way. This procedure may be 
used if only even parity fermionic loops are present \cite{Bonneau}. 
Since the spinor properties remain unchanged, this regularization 
scheme is well suited for supersymmetric calculations \cite{Capper}. 

We return to the situation in 5 dimensions. If the theory has 
5 non-compact dimensions, we can just apply any procedure, 
described above. In (naive) dimensional regularization now the 5 gamma 
matrices are extended to $D$. However, we do not have the difficulty 
with the chirality operator as is trivial  in odd dimensions. On the 
contrary, if the 5th dimension is compactified, the 5 dimensional 
Lorentz invariance is broken, hence $\gg^5$ has a special role. 
This means that we are in a similar situation as in 4 dimensions. 
Therefore, using dimensional reduction the treatment is the same as 
in 4 dimensions. 

We now investigate how dimensional regularization of the Clifford 
algebra can be generalized to included a compact dimension. 
In principle, we can now introduce both additional gamma matrices for 
both the 4 dimensional Minkowski space and the compact manifold. 
However, similar consistency arguments as those presented in 
\cite{Breitenlohner} apply. Depending on whether the additional gamma 
matrices for the compact dimension $\gg^a_\perp$ commute or 
anti-commute with $\gg^5$, we find 
\equ{
(D_5 - 1) \tr\,  \gg^\gm\gg^\gn\gg^\gr\gg^\gs \gg^5 =
\tr\, \gg^a_\perp \gg^a_\perp \gg^\gm\gg^\gn\gg^\gr\gg^\gs \gg^5 = 
\tr\, \gg^a_\perp \gg^\gm\gg^\gn\gg^\gr\gg^\gs \gg^a_\perp \gg^5 = 
\non \\
\pm \tr\,   \gg^a_\perp \gg^\gm\gg^\gn\gg^\gr\gg^\gs \gg^5 \gg^a_\perp = 
\pm (D_5 -1) \tr\,  \gg^\gm\gg^\gn\gg^\gr\gg^\gs \gg^5.
}
In the anti-commuting case (where we pick up the minus) 
we see that for all dimensions $D_5 \neq 1$ 
we find that the trace $\tr\,  \gg^\gm\gg^\gn\gg^\gr\gg^\gs \gg^5$ 
has to vanish. As we certainly want to avoid this, 
we have to make the commuting choice. This means that 
there is no distinction between the additional gamma matrices 
for the 4 non-compact dimensions and the 1 compact dimension, 
so that as far as the gamma algebra in dimensional regularization is 
concerned it goes exactly the same as in the standard 4 dimensional 
case.

\section{Example: an effective one-loop potential}
\labl{EffPot}

A typical form of an effective potential with a KK tower formally reads
\equa{
V_\go = & \ \int \frac{\d^4 p_4}{(2\gp)^4} \sum _{n \in \Intr} \ln \left[
p_4^2 + \Bigl(\frac nR + \go\Bigr)^2 + m^2 
\right] 
\non \\ = &\  
\frac {1}{2\gp i} 
\int_\ominus \d p_5 \int d^4 \frac{\d p_4}{(2\gp)^4} \cP_\go(p_5) 
\ln( p_4^2 + p_5^2 +m^2 ),
}
where $R$ is the compactification radius, $\go$ a 
shift parameter in the lattice of five dimensional momenta. In addition, 
$m$ is a mass parameter that either represents a physical mass or the 
IR regulator introduced to shift the poles away from the real axis, as 
discussed above. This type of mass spectrum $(\frac nR + \go)^2 + m^2$ 
was investigated in \cite{Barbieri,Delgado:98} 
for compactification of the fifth dimension 
on an orbicircle $S^1/\Intr_2$ or $S^1/\Intr_2 \times \Intr_2'$. 

The techniques of manipulation of the pole functions like the one discussed 
in section \ref{sumasint} can be used here as well, therefore we just quote 
results here. The 5 dimensional divergent part of the effective 
potential reads 
\equ{
V_{\go\, 5\,D} = 
- \sin^2  \frac \gp2 D_5\,\frac{\gp^{\half (D_4+D_5)}}{(2\gp)^{D_4}} \, 
\gG( - \shalf (D_4 +D_5)) 
\frac{R\, m^{D_4 + D_5}}{\gm_4^{D_4-4}\gm_5^{D_5-1}}.
\labl{5ddiv}
}
Note that this result is independent of the momentum lattice shift 
parameter $\go$, because the part $\mp i \gp R$ of the pole function 
\eqref{remainder} which is responsible for the $5$ dimensional 
divergence does not depend on $\go$.
In the limit $D_4 \ra 4$ and $D_5 \ra 1$ we find a finite result. 
As explained in our general discussion in section \ref{proppolefunction}, 
this is just an artifact of dimensional regularization.

Apart from this 5 dimensional divergent part, we obtain a finite part 
\equ{
V_{\go\, finite} = - 
\frac {2 \gp^{\half(D_4 + D_5)}}{(2\gp)^{D_4}} \,
\frac{\gG(D_4+D_5)}{\gG(\shalf D_4+ 1) \gG(\half D_5) }\, 
\frac {(2\gp R)^{1-D_4-D_5}}{\gm_4^{D_4-4} \gm_5^{D_5-1}} \, 
K(\go R).
}
Here the poly-logarithm
\(
L_\gs(z) = \sum_{n \geq 1} \frac {z^n}{n^\gs}
\)
 is introduced in the definition of the function 
\equ{
K(\go R) = 
e^{i \frac \gp2(D_5-1)} L_{D_4+D_5}(e^{-2\gp i\, \go R}) + 
e^{-i \frac \gp2(D_5-1)} L_{D_4+D_5}(e^{2\gp i\, \go R}). 
}
We conclude that the effective potential $V_\go$ has a quintic 
divergence and a finite part; a quadratic divergence is absent 
\cite{Ghilencea,GNS}. 
In the model of \cite{Barbieri} the boson and fermion masses are 
related by $\go_F = \go_B + \half \frac 1R$. Then in the effective 
one loop potential the overall 5 dimensional divergent term \eqref{5ddiv} 
cancels and we find the finite contribution 
\equ{
V_{eff} = V_{\go_B\, finite} - V_{\go_F\, finite}.
}

\section{Conclusion}

The main purpose of this paper was to extend the dimensional 
regularization procedure to a space-time with compact dimensions. 
In particular we focused on $M^4 \times C$, with $C$ a compact 
one dimensional manifold.  Since the topology and geometry 
of the 5th dimension is different from the 4 dimensional Minkowski 
space, two regulators may in principle be needed to regularize 
such a theory. 
From experience with un-compact dimensions 
we have learned that dimensional regularization is a powerful 
procedure. However, it cannot be applied directly to a compact dimension 
because we are confronted with sums rather than integrals over the 
momentum. Using complex function theory and a 
``pole function'' (that has poles at the KK momenta), the sum over the 
KK tower can be turned into a contour integral, which may be extended 
to an arbitrary complex dimension in a manner inspired by dimensional 
regularization. In this way we have constructed a regularization 
prescription that carefully treats the additional dimension 
separately, just as it was done in ref.\ \cite{Ghilencea}, 
without  any prejudice of how the regulators should be related. 
But at the same time our regularization prescription reflects 
the properties of the KK towers since they are encoded in a 
regularization independent pole function.
The latter was not respected by putting a momentum 
cut-off on the KK sums in their prescription, unless the 
cut-off for the sum was taken much larger than the cut-off 
for the 4 dimensional momentum. 

The extension of this method so as that fermion loops can 
also be treated. Just like in 4 dimensions, there are various 
prescriptions possible. We discussed dimensional reduction 
where the gamma algebra and the traces are performed in 5 dimensions 
and after that the momentum sum and integrals are regularized 
using dimensional regularization. Secondly, we discussed dimensional 
regularization of the Clifford algebra. In order not to run into an 
inconsistency, it was necessary that the additional gamma matrices 
for the compact dimension commute with $\gg^5$ just like the
additional gamma matrices in 4 dimensions. 

Using the dimensional regularization procedure of both compact 
and non-compact dimensions, the possible types of
divergences can be identified: there can be true 5 dimensional divergences,  
4 dimensional divergences and finite contributions. This is encoded in 
the asymptotic behavior of the pole function: for the orbifold 
$S^1/\Intr_2$, for example, the pole function has a 
constant, a single pole and an exponentially suppressed part for the 
5 dimensional momentum going to $i \infty$; these correspond to the 
afore mentioned 5, 4 dimensional divergent and finite parts. 
Because this method makes transparent the structure of divergences 
that can occur and 
what the nature of these are; it also easier to understand why they 
may cancel in certain situations. 

As an example of this the effective 
one-loop potential due to Yukawa interactions in the model of 
ref.\  \cite{Barbieri} was calculated using 
this prescription. For both the bosonic and fermionic parts 
of this potential we found the same 5 dimensional divergent piece but 
different finite parts. The fact that there is an equal number of 
bosonic and fermionic towers (a left over of the $N=2$ supersymmetry 
that was present before the orbifolding) the divergent pieces 
cancel out. We should stress that this one-loop calculation does not 
show that is free of divergences to all orders, or even at one-loop since 
only Yukawa couplings are taken into account.  
Higher order corrections, like the two-loop Yukawa corrections 
to the potential calculated in ref.\ \cite{Delgado} are divergent. 
However, their result seems to show that the superpotential on 
the branes does not renormalize as they are able to absorb the 
linear divergence by wave function renormalization. 
In ref.\  \cite{SHPD} we show that the Fayet-Iliopoulos tadpole can 
be quadratically divergent using methods discussed in the paper.

\section*{Acknowledgements} 

The author thanks H.P.\ Nilles and D.\ Ghilencea for stimulating 
discussions that initiated this work, and Marek Olechowski for 
useful comments.
This work is supported by priority grant 1096 of the Deutsche 
Forschungsgemeinschaft and European Commission RTN 
programmes HPRN-CT-2000-00131 and 
HPRN-CT-2000-00148.

\end{document}